\documentstyle[twocolumn,prb,aps]{revtex}
\begin{document}
\draft
\title{ Low temperature thermodynamics of charged bosons in a random 
potential and the specific heat of $La_{2-x}Sr_{x}CuO_{4}$ below $T{c}$.}
\author{A.S. Alexandrov \and R.T. Giles}
\address{Department of Physics, Loughborough University, Loughborough, LE11
3TU, U.K.}
\date{\today}
\maketitle

\begin{abstract}
We propose
 a simple 
analytical form of the partition function for
charged bosons localised in a random
potential and derive the consequent thermodynamics
below the superfluid transition temperature.
In the low temperature limit, the specific heat
$C$ depends on  the localisation length exponent
$\nu$: $C$ is linear for $\nu\leq1$, but for $\nu>1$ we obtain
$C\propto T^{1/\nu}$.
This unusual sub-linear temperature dependence of
the specific heat has recently been observed
in $La_{2-x}Sr_{x}CuO_{4}$ below 
$T{c}$.   
\end{abstract}


\section{Introduction}

A charged $2e$ Bose liquid of small bipolarons
  as a microscopic 
model of the ground state of cuprates \cite{alemot} explains 
the high value of $T_{c}$ and its `boomerang' 
doping dependence \cite{ale3},  the divergent upper critical 
field \cite{ale2},  
 the normal state in-plane and `c'-axis 
transport, 
the magnetic susceptibility and  
the pseudogap\cite{alekab}. The (bi)polaronic nature of 
carriers in these materials is supported by  observations 
of the characteristic polaronic spectral function in the infrared 
spectra \cite{mic}, of the isotope effect on the carrier 
mass\cite{mul}, of  flat bands in the high-resolution photoemission 
spectra\cite{kin}, and  by a few other experiments such as the 
pair-distribution analysis of neutron scattering \cite{sen}. 

The bipolaron model overcomes  a  fundamental problem with any theory 
of high-$T_{c}$ which is  the short-range  
  Coulomb repulsion. As stressed by 
  Emery $et$ $al$ \cite{em} the theories involving real-space pairs with a pure electronic 
  (exchange)
  mechanism of pairing are $a$ $priori$ implausible due 
  to the strong short-range Coulomb repulsion between two carriers. 
  The 
  direct  repulsion is usually 
   much stronger than any exchange interaction. 
   The attractive potential generated by the electron-phonon 
  interaction of the Holstein model may  overcome the short-range 
  Coulomb repulsion, but inevitably involves  a huge carrier mass enhancement 
  (otherwise the phonon frequency would be extremely high).  
 On the other hand, one of us  showed \cite{ale3} that the Fr\"{o}hlich 
 electron-phonon interaction can provide
  $intrinsically$ mobile intersite small bipolarons which  
    are condensed at  high $T_{c}$ (of the order of $100K$). 
   This interaction, operating on a scale of the order of $1 
    eV$,   compensates the Coulomb repulsion and allows 
    the deformation potential (together with an exchange interaction of 
    any origin) to bind two holes into an intersite mobile bipolaron in the 
    $CuO_{2}$ plane.  The mass 
    renormalisation appears to be 
    smaller by several orders of magnitude than in the 
    Holstein model (with the same attractive potential). 
   Hence the charge $2e$  Bose liquid, being already an important 
reference system of quantum statistics, is
 now  of particular physical
interest. Because doped Mott insulators are  disordered, the 
localisation of 
carriers by a random potential plays an  important role in their 
low-temperature thermodynamics and transport  \cite{alemot}.
 
 In this paper we study the  thermodynamics of  charged bosons 
in the 
 superfluid phase assuming that some of them are localised by disorder. 
 The fact that in the superconducting phase the chemical potential 
$\mu$ is pinned at the mobility edge, $\mu=E_{c}$,  significantly simplifies 
the problem. By using a plausible form of the single well 
partition function and a model density of localised states, we find the 
temperature dependence of the specific heat.   Under certain 
conditions the predicted specific heat follows precisely the sub-linear
temperature dependence observed\cite{nai} in
$La_{2-x}Sr_{x}CuO_{4}$  at  low temperatures.

\section{Partition function}

 The  picture of interacting bosons with a short-range 
interaction 
filling up all the localised single-particle states in a random potential 
 is  known in the 
literature \cite{her,ma,fis}.
To calculate the partition function of localised bosons  one has to
 take into account  the repulsion 
between them.  One cannot ignore the fact that the
localisation length $\xi$ generally varies with energy and diverges at 
the mobility edge. One would expect that the number of $hard$ $core$ 
bosons 
in a localised state near the mobility edge diverges in a similar way to
the localisation length. Only a repulsive interaction can
stop all particles condensing into the lowest localised state. Thus, as 
stressed by Fisher $et$ $al$ \cite{fis}, for a Bose gas (in a random field)
 there is no sensible
non-interacting starting point about
which to perturb; this is in
contrast to the case of a Fermi gas.

 Although the comprehensive scaling analysis of neutral \cite{fis} 
and  charged
bosons  \cite{fis2} allows us to describe  
the quantum Bose glass-superfluid transition, the shape of the scaling 
functions as well as  the thermodynamics of each phase away from the 
transition remain 
 unknown. A physically plausible model of neutral \cite{gun} 
and charged bosons \cite{alegil} in a random potential based on a separation of 
localised single-particle states from delocalised states might be 
 helpful.
As discussed in Ref.\cite{fis2},   such separation, strictly speaking, is   not 
possible as the localised states will be subsumed into the collective 
mode. Nevertheless, the effect of this mode can be  understood 
by analysing the high density limit where the 
dimensionless Coulomb repulsion is small: $r_{s}=4me^{2}/\epsilon_{0}
(4\pi n/3)^{1/3}\leq 1$. Here $m$ is the boson mass, $n$ is the 
density and $\epsilon_{0}$ the 
static dielectric constant of the host material (and we take $\hbar=1$).
In this limit the
excitation
spectrum at $T=0$ was calculated by Foldy \cite{fol}, who worked at zero 
temperature using the Bogoliubov \cite{bog} 
approach. The Bogoliubov method leads to the result that 
 the elementary excitations of the system have, for small 
momenta, energies characteristic of plasma oscillations which pass over 
smoothly for large momenta to the energies characteristic of single 
particle excitations. They screen an external charge with a 
 screening radius, which  
is temperature independent at low temperatures
 \cite{hor,fra}. At high density, $r_{s}\ll 1$, one can expect a 
disappearence of localised states because of the screening  similar to the 
Mott metal-insulator transition in doped semiconductors.

It is more likely, however, that   the  dimensionless strength of the
Coulomb repulsion is about unity or larger, 
$r_{s}\geq 1$, and it is only when $r_s$ is very large $r_{s}>>>1$ 
that the Wigner crystallisation of charged carriers occurs. Hence, 
there is  a  wide interval of densities where the localised  states coexist 
with a superfluid. Being screened they depend on the interaction and 
the superfluid density. Nevertheless, based on the screening in the high density limit 
\cite{hor,fra} we expect that the density of localised states
$\rho_{L}$
near the  mobility edge, $E_{c}$ remains  temperature independent at low 
temperatures ($k_B T\rho_{L}\ll 1$). If this is so, then the  renormalisation of the 
effective single-particle energies  by the collective mode does not 
affect the temperature dependence of any of the thermodynamic 
functions at low temperatures.

With all these reservations 
we assume that  at some 
 finite temperature $T_{c}$, bosons are  condensed at $E=E_{c}$ so that 
the chemical potential $\mu=E_{c}$. 
 The excitation spectrum of the delocalised  superfluid has a gap of 
the order 
of the plasma frequency \cite{fol}
and so the Bogoliubov collective modes can be ignored
in the thermodynamics of the system (in $3D$
their contribution is exponentially small while in $2D$ their energy 
scales as $T^{5}$ and the specific heat as $T^{4}$).
Even in the
case of a short range repulsion the sound modes yield an energy 
proportional to $T^{d+1}$  and hence a specific heat which behaves 
like $C\propto T^{d}$ (where $d = \mbox{dimensionality}$) \cite{fis}.
Thus,  for $d\geq 2$,
the contribution to thermodynamics from the delocalised bosons appears to be 
negligible at low temperatures compared with that from bosons localised in 
shallow potential wells. Hence in the following we 
calculate the partition function and specific heat of  localised 
bosons only.

For simplicity, we choose
\begin{equation}
   E_c=0
 \end{equation}
When two or more charged bosons
are in a single localised state of energy $E$ there may be significant Coulomb energy 
and we try to take this into account as follows.  The localisation length 
$\xi$ is thought to depend on $E$ via
\begin{equation}
  \xi\propto \frac{1}{\left(-E\right)^\nu}
  \label{eq:xi}
  \end{equation}
  where $\nu>0$. The Coulomb potential energy of $p$ charged bosons confined within a 
radius $\xi$ can be expected to be
\begin{equation}
   \mbox{potential energy }\sim\frac{p(p-1)e^2}{\epsilon_0\xi}.
   \label{eq:pe}
\end{equation}
Thus the total energy of $p$ bosons in a localised state of energy 
$E$ is taken to be
\begin{equation}
  w(E)=pE + p(p-1)\kappa\left(-E\right)^\nu
  \label{eq:w}
\end{equation}
where $\kappa>0$. Hence we see that the behaviour of charged bosons in 
localised states can be thought of as intermediate between Bose-Einstein 
statistics and Fermi-Dirac statistics.  When $\kappa=0$ we have an equally 
spaced set of levels, i.e. Bose-Einstein behaviour, whereas when 
$\kappa=\infty$ we have Fermi-Dirac behaviour since the only levels with 
finite energy are $p=0$ and $p=1$, thus enforcing an exclusion principle.
When $0<\kappa<\infty$ we have the intermediate `parastatistics' that 
the level spacing $\delta w\rightarrow\infty$ as $p\rightarrow\infty$.

We take the total energy of  charged bosons in localised 
 states to be the sum of the energies of the bosons in the 
individual potential wells.  The partition function $Z$ for such a system is then 
the product of the partition functions for each of the wells,
 and the system free energy $F=k_B T \ln Z$ is simply the 
sum of the individual free energies $k_B T\ln z(E)$.
The free energy of 
the localised bosons in one unit cell is then given by
\begin{equation}
F=k_B T\int_{-\infty}^{0}dE \rho_{L}(E)\ln z(E),
\label{eq:F}
\end{equation}
where $\rho_{L}(E)$ is the one-particle density of localised states
per unit cell below the mobility edge.

In this paper we assume the following physically plausible
density  of localised states  $\rho_{L}(E)$:
\begin{equation}
  \rho_{L}(E) = \frac{n_L}{\gamma}e^{\frac{E}{\gamma}}
  \label{eq:rho}
\end{equation}
We shall see below that, in the most interesting case, if the width
of the tail $\gamma$ is large compared
with $k_B T$, the
 specific heat is 
insensitive  to the details of the shape of $\rho_L(E)$, depending only
on the value of $\rho_L(0)$.

\section{The thermodynamics of a single potential well}

Before considering the thermodynamics of charged bosons in our assumed
density of localised states,
we first summarise here the results we have previously obtained\cite{alegil}
for the thermodynamics of a single potential well with only one single
particle level of energy $\epsilon$, i.e. for the case
\begin{equation}
  \rho_L(E)=n_L\delta(E-\epsilon)
\end{equation}
The
probability for this state to be occupied by $p$ bosons is proportional to
\[ e^{-\beta \left\{w(\epsilon)-p\mu\right\}} \]
where $\mu$ is the chemical potential and $\beta \equiv 1/k_{B}T$. 
We can re-express $w-p\mu$ as 
\begin{equation}
 w-p\mu = 
         \kappa\left(-\epsilon\right)^\nu\left(p-p_0\right)^2
    		-\kappa\left(-\epsilon\right)^{\nu}p_{0}^{2}
\end{equation}
where
   \begin{equation}
      p_0 =  \frac{1}{2} + \frac{\mu-\epsilon}{2\kappa(-\epsilon)^\nu}
   \end{equation}
Fig 1 shows a graph of $w-p\mu$ as a function of $p$.  The partition 
function $z(\epsilon)$ for such a single localised state is
\begin{eqnarray}
    z(\epsilon) & = & \sum_{p=0}^{\infty} e^{-\beta (w-p\mu)}  \nonumber \\
            & = & e^{p_{0}^{2}\beta \kappa\left(- \epsilon\right)^{\nu}}
       \sum_{p=0}^{\infty}
	e^{-\beta \kappa\left(-\epsilon\right)^{\nu}\left(p-p_0\right)^2}.
		\label{e:zls}
\end{eqnarray}
The partition function is thus completely determined by the dimensionless 
parameters $p_0$ and $k_{B}T/[\kappa(-\epsilon)^{\nu}]$.

The corresponding mean occupancy $\left<p\right>$ given by
\begin{equation}
	\langle p \rangle=k_{B}T\frac{\partial \ln  z(\epsilon)}{\partial\mu}  \label{e:<p>}
	\end{equation}
and specific heat capacity $c$ given by
\begin{equation}
	c = \beta^{2}\frac{\partial^2 \ln  z(\epsilon)}{\partial \beta^{2}}.
			\label{e:c_v}
	\end{equation}
are shown in Figs~2 and~3 for the case $\mu=0$.  These results were
calculated by simply truncating the partition function series at 100 terms.

We now attempt to understand these 
results in more detail, looking separately at each temperature range.

\begin{enumerate}
  \item $k_{B}T\ll\kappa\left(-\epsilon\right)^{\nu}$
  
At low temperatures the partition function is dominated by the term with 
$p$ closest to $p_0$, i.e. the value of $p$ giving the lowest value of 
$w-p\mu$, and so the mean occupancy $\langle p \rangle$ is an integer and 
goes up in steps as $p_0$ increases, as seen in Fig~2.  The changeover in 
dominance from one term to another occurs when $p_0$ is a half-integer, at 
which point the two lowest energy states are degenerate.

So long as one term dominates the partition function, the specific heat 
 $c$ will be close to zero.  However when $p_0$ is close to a 
half-integer we have a two level system and a corresponding Schottky 
anomaly in the specific heat capacity.    Hence, at fixed 
temperature,  the low temperature specific heat capacity (i.e. $k_{B}T\ll p_0^2 
\kappa(-\epsilon)^\nu$) is periodic in $p_0$. 

  \item $k_{B}T>\kappa(-\epsilon)^{\nu}$ 
  
  We can approximate the sum by an integral
\begin{equation}
 z(\epsilon) \approx  
e^{p_{0}^{2}\beta\kappa\left(-\epsilon\right)^{\nu}}
\int_{0}^{\infty}dp e^{-\beta \kappa(-\epsilon)^{\nu}(p-p_0)^2}
\end{equation}

    \begin{itemize}
      \item $\kappa(-\epsilon)^{\nu}<k_{B}T<p_{0}^{2}\kappa(-\epsilon)^{\nu}$

In this case we can approximate the lower limit of the integral 
as~$-\infty$, i.e. the partition function can be approximated by an 
untruncated gaussian, and is therefore approximately symmetrical about 
$p_0$. Hence, in this temperature range we have
  \begin{equation}
      \langle p \rangle \approx p_0
  \end{equation}
as is clearly seen in Fig~2, and
\begin{equation}
  c\approx\frac{1}{2}k_B
\end{equation}
as seen in Fig~3.

\item $k_{B}T>\kappa(-\epsilon)^{\nu}$ and $k_{B}T \gg 
   p_{0}^{2}\kappa(-\epsilon)^{\nu}$

In this case we can approximate the partition function as being
an integral over half a Gaussian.  We then obtain
   \begin{equation}
	\langle p \rangle \approx \sqrt{\frac{k_{B}T}{\pi\kappa(-\epsilon)^\nu}}
	\end{equation}
and the specific heat is again given by $c\approx k_B/2$; both of these
results are seen in Figs~2 and~3.
 
 \end{itemize}
    \end{enumerate}

\section{The number of bosons in localised states}

Having established the partition function $z(E)$ for a single
potential well containing one single particle level
of energy $E$ we now use equations~\ref{eq:F}
and~\ref{eq:rho} to derive the thermodynamics of localised bosons.

The average number $N_L$ of bosons in localised states in each unit cell is
\begin{equation}
  N_L = \int_{-\infty}^{0} dE \left<p\right> \rho_L(E)
\end{equation}
 $\left<p\right>$ can be approximated as
\begin{equation}  
  \left<p\right>\approx \left\{ \begin{array}{ll}
      p_o &  \mbox{if $k_BT<p_o^2\kappa(-E)^{\nu}$} \\
      \sqrt{\frac{k_B T}{\pi\kappa(-E)^{\nu}}}
          &  \mbox{otherwise}
      \end{array} \right.
\end{equation}
Hence a reasonable approximation for $\left<p\right>$ at all temperatures is
\begin{equation}
  \left<p\right>\approx \frac{1}{2}
          + \frac{1}{2\kappa(-E)^{\nu-1}}
          + \sqrt{\frac{k_B T}{\pi\kappa(-E)^{\nu}}}
\end{equation}
With this approximation we obtain
\begin{equation}
N_L = n_L \left\{ \frac{1}{2} + \frac{\Gamma(2-\nu)}{2\kappa\gamma^{\nu-1}}
       + \sqrt{\frac{k_B T}{\pi\kappa\gamma^{\nu}}}\Gamma(1-\frac{\nu}{2})
       \right\}
\end{equation}
Hence if $\nu>2$ then $N_L$ becomes infinite and so the formation
of a superfluid is excluded.

\section{The specific heat of localised bosons}
We have shown above that the contribution to the  specific heat $c$ 
from a single localised state 
is determined solely by
the values of two dimensionless quantities: $p_o$ and
$k_BT/\left[\kappa(-E)^\nu\right]$.  If $\nu$ is
temperature independent,
 $c$ can instead
be considered as a function of a different pair of quantities:
$-E/\left[k_BT\right]$ and $\tau=\kappa\left(k_BT\right)^{\nu-1}$
The integrated specific heat C of the localised bosons is
then
\begin{equation}
  C = \int_{-\infty}^{0}dE\,\rho_L(E)
      c\left(\frac{-E}{k_BT},\tau\right)
  \label{eq:C}
\end{equation}

\subsection{Specific heat for $\nu=1$}
In this case $\tau=\kappa$ and is therefore temperature independent.
Thus we obtain:
\begin{equation}
  C = k_B T\int_{-\infty}^{0}d\left(\frac{-E}{k_BT}\right) \,\rho_L(E)
      c\left(\frac{-E}{k_BT},\kappa\right)
      \label{eq:nu=1}
\end{equation}
If $\rho_L(E)$ is constant in the region $-E<\sim k_BT/\kappa$ then
equation~\ref{eq:nu=1} yields $C\propto T$; for $\rho_L$ given by
equation~\ref{eq:rho} this result applies if $\gamma\gg k_BT/\kappa$.

\subsection{Specific heat for $\nu>1$}
For $\nu>1$ the integral in equation~\ref{eq:C} can be thought of as
an integral over three regions:
\begin{enumerate}
\item $-E/\left[k_BT\right]<\tau^{-1/\nu}$
      \ \{i.e. $p_o>\frac{1}{2}\left(1+\tau^{-1/\nu}\right)$~\}

      In this region $k_BT>\kappa(-E)^\nu$ and so
      \begin{equation}
           c\approx\frac{1}{2}k_B
           \label{eq:c1}
      \end{equation}
\item $\tau^{-1/\nu}<-E/\left[k_BT\right]<\tau^{-1/(\nu-1)}$
      \begin{flushright}
        \{i.e. $\frac{1}{2}\left(1+\tau^{-1/\nu}\right)>p_o>1$~\}
      \end{flushright}

      This region only exists if $\tau<1$.
      $c$ is dominated by Schottky anomalies, each centred on a half-integer
      value of $p_o$.  The last full Schottky anomaly in the series is at
      $p_o=\frac{3}{2}$, so the total number $s$ of Schottky anomalies is
      \begin{equation}
         s=\frac{1}{2}\tau^{-1/\nu}-\frac{1}{2}
         \label{eq:s}
      \end{equation}
      If we ignore all except the lowest two energy levels then
      for each anomaly
      \begin{equation}
        \int_{below}^{above}dE\,
           c\left(\frac{-E}{k_BT},\tau\right)
               = \frac{\pi^2}{3(\nu-1)}k_B^2T
        \label{eq:c2}
      \end{equation}
\item $-E/\left[k_BT\right]>\tau^{-1/(\nu-1)}$ \ \{i.e. $p_o<1$~\}

      Here the $p=1$ level is the lowest energy level, and the $p=0$ level is
      the next lowest.  If higher levels are neglected then $c$ is solely a function
      of $-E/\left[k_BT\right]$ and is non-negligible only for
      $0.1<-E/\left[k_BT\right]<10$.  Hence this region can be
      neglected entirely if $\tau<\left(\frac{1}{10}\right)^{\nu-1}$.  If, on
      the other hand, $\tau>10^{\nu-1}$ the integrated specific heat for this region
      will be that of half a Schottky anomaly:
      \begin{equation}
        \int_{region\,3}\!\!\!\! dE\,
          c\left(\frac{-E}{k_BT},\tau\right)
            \approx \frac{1}{6}\pi^2k_B^2T
        \label{eq:c3}
      \end{equation}
\end{enumerate}

If $\kappa\gamma^{\nu-1}\gg1$ then $e^{E/\gamma}\approx 1$
throughout regions
(1) and (2).  If $\gamma\gg k_BT$ then $e^{E/\gamma}\approx 1$
throughout region (3).  We take both of these conditions to hold so that
equation~\ref{eq:C} can be re-written as
\begin{equation}
  \frac{C}{T} \approx \frac{n_Lk_B}{\gamma}\int_{0}^{\infty}
      d\left(\frac{-E}{k_BT}\right)\,
      c\left(\frac{-E}{k_BT},\tau\right)
  \label{eq:C/T}
\end{equation}
i.e. if $\nu$ is fixed then $C/T$ is solely a function of $\tau$.

Hence we can distinguish two cases for which an approximate analytical formula
for $C$ can be derived:

\begin{enumerate}
\item $\tau<\left(\frac{1}{10}\right)^{\nu-1}$

     In this case region (3) can be neglected. From equations~\ref{eq:c1},
     \ref{eq:s} and~\ref{eq:c2} we then obtain
     \begin{equation}
       \frac{C}{T} \approx \left[\frac{\pi^2}{6(\nu-1)}+\frac{1}{2}\right]
                     \frac{n_Lk_B^2}{\tau^{1/\nu}\gamma}
        \label{eq:C/T1}
     \end{equation}
     This implies
     \begin{equation}
       C \propto T^{1/\nu}
     \end{equation}
     i.e. the specific heat has a power law dependence on temperature in
     which the power is less than unity.
\item $\tau>10^{\nu-1}$

     In this case the specific heat $C$ is dominated by region (3).  From
     equation~\ref{eq:c3} we then obtain
     \begin{equation}
        \frac{C}{T} \approx \frac{\pi^2n_Lk_B^2}{6\gamma}
        \label{eq:C/T2}
     \end{equation}
     i.e. $C/T$ is temperature independent.
\end{enumerate}

Fig.~4 compares these two formulae (for the case $\nu=1.5$)
 with the result of calculating $C$
from equation~\ref{eq:C/T} by numerical integration (in which $c$ is calculated
from the first 100 terms of the partition function).

\subsection{Specific heat for $\nu<1$}

Now $p_o$ increases as $-E$ increases.
Again we can distinguish three regions
\begin{enumerate}
\item $-E/\left[k_BT\right]<\tau^{-1/\nu}$
      \ \{i.e. $p_o<\frac{1}{2}\left(1+\tau^{-1/\nu}\right)$~\}

      Here $k_BT>\kappa(-E)^\nu$ and so once again $c\approx k_B/2$.
\item $\tau^{-1/\nu}<-E/\left[k_BT\right]<\tau^{1/(1-\nu)}$
      \begin{flushright}
        \ \{i.e. $\frac{1}{2}\left(1+\tau^{-1/\nu}\right)<p_o<1$~\}
      \end{flushright}

      Here the partition function is dominated by the two levels corresponding to
      $p=1$ and $p=0$.  If $\tau<1$ this region does not exist.

\item $-E/\left[k_BT\right]>\tau^{1/(1-\nu)}$ \ \{i.e. $p_o<1$~\}

     Here $c$ is dominated by an infinite series of Schottky anomalies
     centred on values of $E$ corresponding to half-integer values
     of $p_0$:
     \begin{equation}
       -E=\left[(2p_0-1)\kappa\right]^\frac{1}{1-\nu}
     \end{equation}
       
\end{enumerate}

We consider the low temperature limit, i.e. $\tau\gg1$ (for $\nu<1$).
Let
\begin{equation}
  \chi = \frac{-E}{\gamma}
\end{equation}
Equation~\ref{eq:C} becomes
\begin{equation}
  C=n_L\int_0^\infty e^{-\chi} c\,d\chi
  \label{eq:Cchi}
\end{equation}
In the low temperature limit $k_BT\ll\kappa\gamma^\nu$, region~1 makes a
negligible contribution to the integral in equation~\ref{eq:Cchi}.
Provided $k_BT\ll(1-\nu)\gamma$ we can make the
approximation that, when integrating through each Schottky anomaly,
$e^{-\chi}$ can be taken to be a constant depending on the anomaly;
hence the contribution from each anomaly is proportional to the result
given in equation~\ref{eq:c2} (and equation~\ref{eq:c3}).  Thus, in
the low temperature limit, we have $C\propto T$.

If $\gamma^{1-\nu}\gg\kappa$ then many Schottky anomalies contribute
significantly to the integral in equation~\ref{eq:Cchi}.  In that case
we can neglect region~2 and transform the sum over Schottky anomalies
into an integral with respect to $p_0$, obtaining:

\begin{eqnarray}
 C    & \approx &  \frac{\pi^2n_Lk_B^2T}{3(\nu-1)\gamma}
                    \int_{0}^{\infty} e^{-\chi} dp_0
        \nonumber \\
    & \approx &  \frac{\pi^2n_Lk_B^2T}{6\kappa\gamma^\nu}
       \int_{0}^\infty \chi^{-\nu} e^{-\chi}d\chi    \nonumber \\
    & \approx & \frac{\pi^2k_B^2n_L\Gamma(1-\nu)}{6\kappa\gamma^\nu}T
\end{eqnarray}

In the high temperature limit, on the other hand, we have $\tau\ll 1$ and 
 $k_BT\gg\kappa\gamma^\nu$; in that case only region~1 is significant and
 we obtain:
\begin{equation}
  C \approx \frac{1}{2}n_Lk_B
\end{equation}

\section{Conclusion}

We have used a 
reasonable scaling of Coulomb energy with localisation length 
to calculate  
the low temperature dependence of the specific heat of a
charged 
superfluid in a random potential. The result strongly depends on the 
exponent $\nu$ of the localisation length. While the 
specific heat is linear for $\nu \leq 1$, it is proportional to
$T^{{1/\nu}}$ for $\nu >1$.

We believe that our findings are relevent for those doped high-$T_{c}$ cuprates 
having many properties  of a  charged 
Bose-liquid\cite{alemot}. Because of the high level of doping of these 
Mott-Hubbard insulators in the superconducting region  one can 
expect a continuous density of states $\rho_{L}(E)$ similar to that studied
by us; our prediction about the power law dependence of $C$
(for $\nu>1$) requires, of the density of states, only that
$\rho_{L}(E_c)\ne 0$.

Superconducting $La_{2-x}Sr_{x}CuO_{4}$ has been observed\cite{nai}
to have a low temperature specific heat proportional to
$T^{\alpha}$ with $\alpha$ being in the range $0.33<\alpha<0.78$
dependent on doping.
This fits with our prediction
for charged bosons partly localised by disorder at low temperature and
$\nu>1$.
The
doping dependence of $\alpha=1/\nu$ could be due to a gradual 
change of the localisation length exponent $\nu$ caused by a difference
in the random potential
profiles for different $x$.

Values of $\alpha$ as low as 0.33 are observed~\cite{nai}
experimentally while
our model excludes the possibility of superconductivity for
$\alpha=1/\nu\le0.5$. However we believe that the approximation for
Coulomb potential energy in
equations~\ref{eq:pe}-\ref{eq:w} is likely to be inadequate
at large~$\nu$.

\vspace{7mm}
  Enlightening   discussions with N. Hussey,
 V. Kabanov,  F. Kusmartsev,  J. Samson, and K.R.A. Ziebeck are 
highly appreciated.

\pagebreak
\begin{figure}
\caption{Graph of $(w-p\mu)/[\kappa(-E)^\nu]$ against $p$.}
\label{fig1}
\end{figure}

\begin{figure}
\caption{The mean occupancy $\langle p\rangle$ of a single level
as a function of $p_0$ and
$\log_{10}\left\{k_{B}T/[\kappa(-\epsilon)^\nu]\right\}$.}
\label{fig2}
\end{figure}

\begin{figure}
\caption{The specific heat capacity $c$ of a single level
as a function of $p_0$ and
$\log_{10}\left\{k_{B}T/[\kappa(-\epsilon)^\nu]\right\}$.}
\label{fig3}
\end{figure}

\begin{figure}
\caption{Specific heat capacity~C divided by temperature~T for the case
$\nu=1.5$.  The continuous curve is obtained from numerical integration
of the integral in equation~\ref{eq:C/T},
$c$ being evaluated by direct summation of the first 100 terms of the
partition function. The two dashed lines are the predictions from
equations~\ref{eq:C/T1} and~\ref{eq:C/T2}.}
\label{fig4}
\end{figure}

 
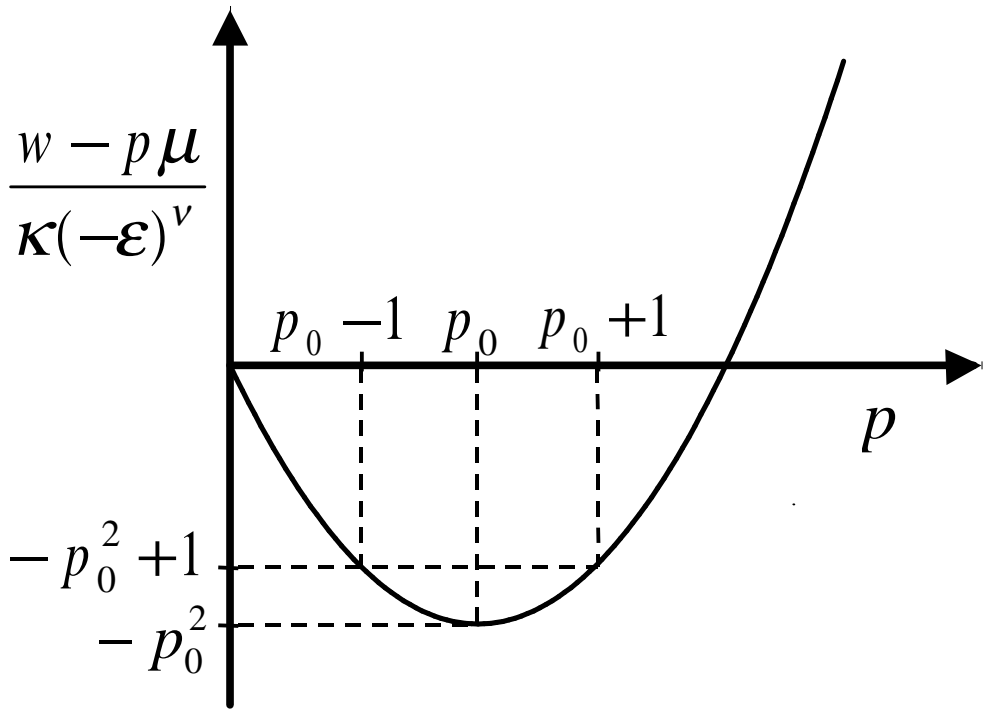
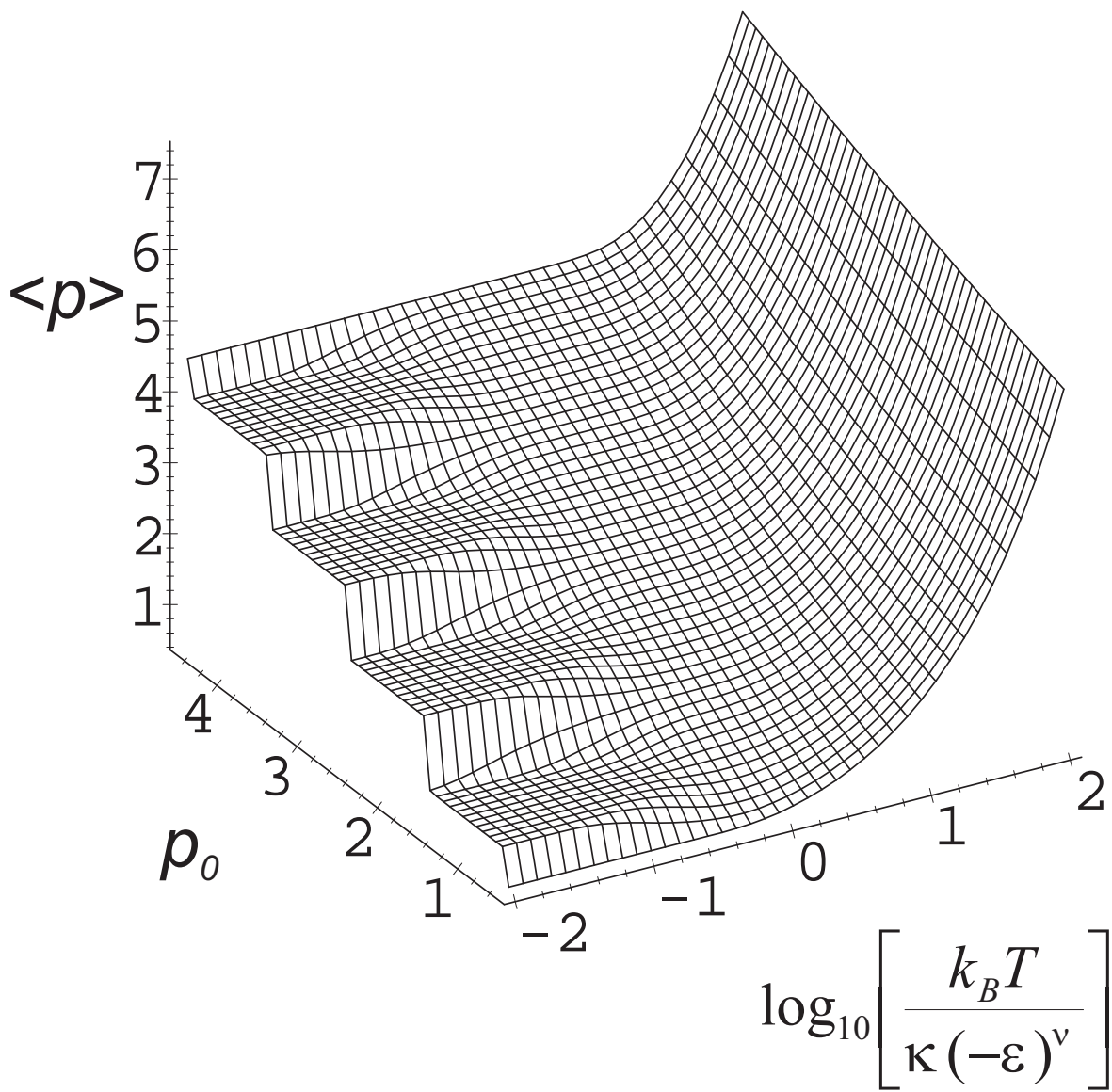
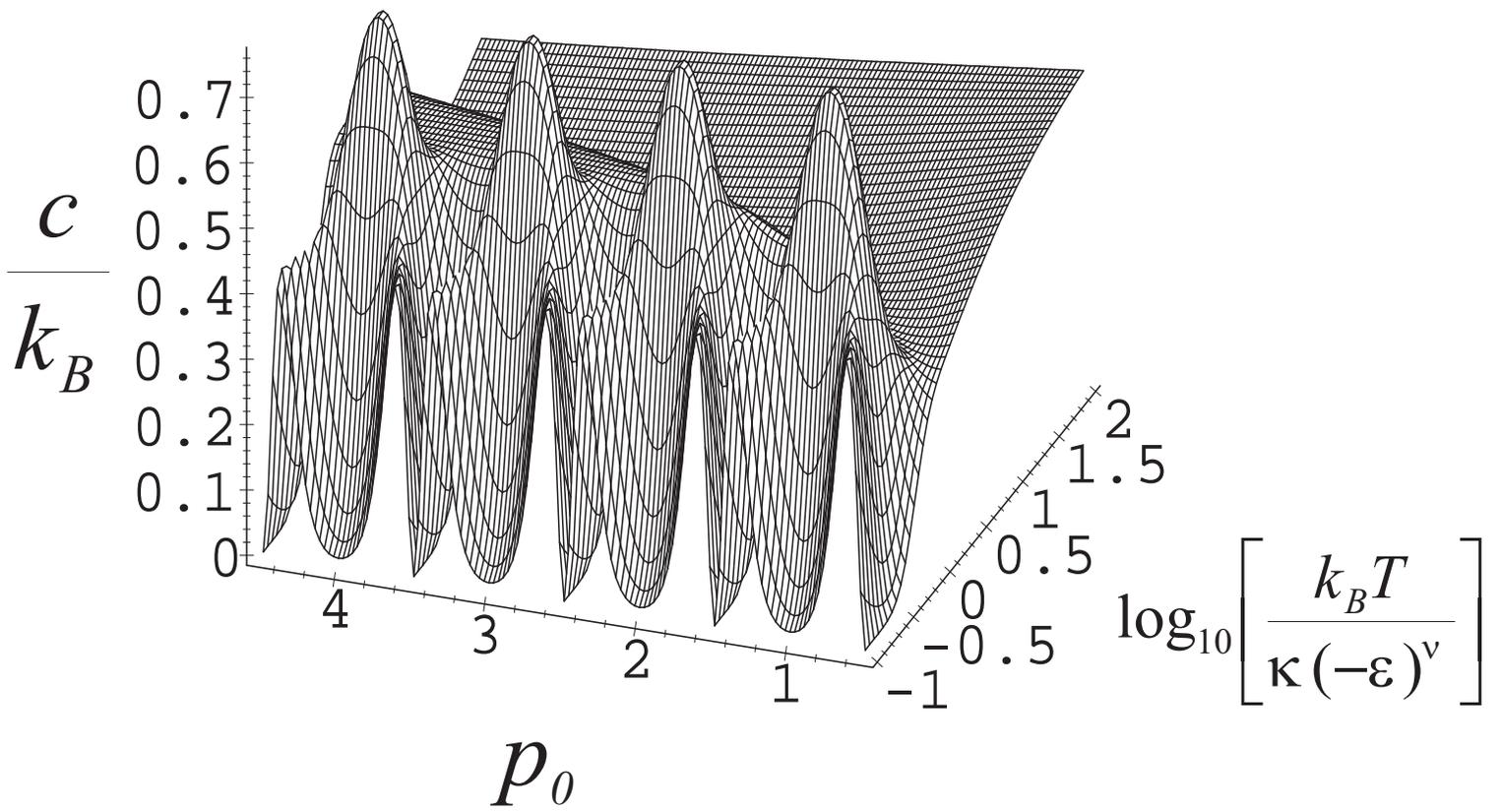
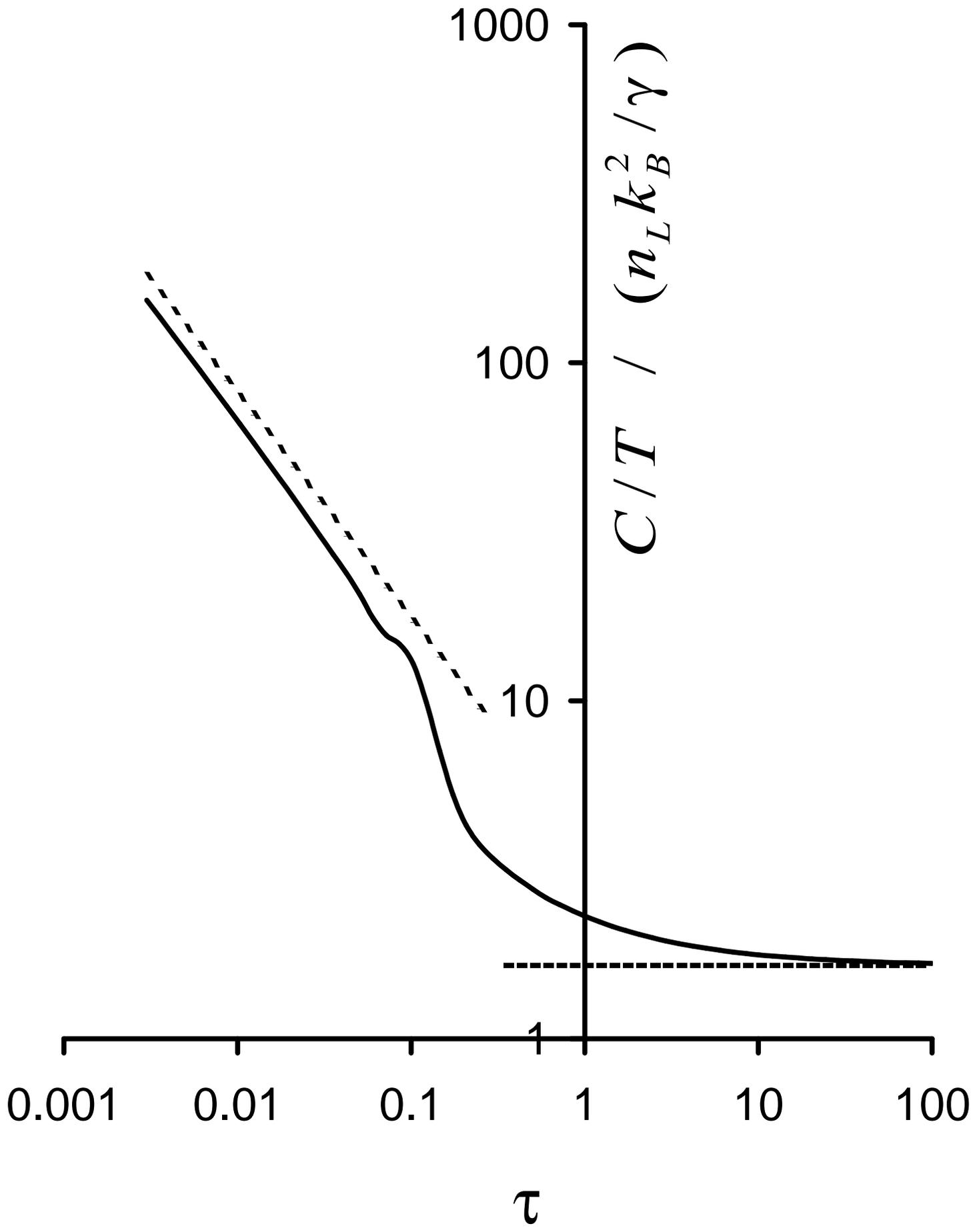
\end{document}